\documentclass[11pt,a4paper]{article}
\usepackage{a4wide}
\usepackage{epsfig}
\usepackage{graphicx}
\usepackage{amsmath}
\usepackage{amssymb}
\usepackage{amsfonts}
\usepackage{hyperref}
\usepackage[chatter]{rotating}
\usepackage{fancyhdr}
\usepackage{makeidx, index}

\newcommand{\sss}[1]{\scriptscriptstyle #1}
\newcommand{\ddx}[1]{\frac{ \delta }{ \delta #1 }}
\newcommand{\dd}[1]{\delta_{#1}}
\newcommand{\wtd}[1]{\widetilde{#1}}
\newcommand{\eqdef}{\stackrel{\rm def}{=}}
\newcommand{\olra}[1]{\overleftrightarrow{\dd{ #1}}}
\newcommand{\ola}[1]{\overleftarrow{\dd{ #1}}}
\newcommand{\ora}[1]{\overrightarrow{\dd{ #1}}}
\newcommand{\dds}[1]{\frac{\overrightarrow{\delta}}{\delta #1 }}

\begin{document}


\begin{center}
{\LARGE The equivalence theorem in effective theories}
\end{center}

\begin{center}
D.~Chicherin$^{1,2,3,\alpha}$,
V.~Gorbenko$^{1,4,\beta}$,
and
V.~Vereshagin$^{1,\gamma}$
\end{center}

\begin{center}
${^\alpha}$tchitcherin@gmail.com \\
${^\beta}$vg629@nyu.edu \\
$^\gamma$vvv@av2467.spb.edu
\end{center}

\begin{center}
\em
$^1$Theor.\ Phys.\ Dept., Institute of Physics,
    St.-Petersburg State University,
    St.-Petersburg, Petrodvoretz,
    198504, Russia \\
$^2$Chebyshev Laboratory, Dept. of Mathematics and
Mechanics, St.Petersburg State University, St.-Petersburg,
199178, Russia  \\
$^3$St.-Petersburg Dept. of Steklov Mathematical Institute,
Russian Academy of Sciences, St.-Petersburg, Fontanka 27,
St.-Petersburg, 191023. Russia  \\
$^4$Center for Cosmology and Particle Physics, Department of Physics,
    New York University, New York, NY, 10003, USA \\
\end{center}

PACS numbers: 03.70.+k, 11.10.-z, 11.55.Ds, 11.90.+t

\begin{abstract}
The famous equivalence theorem is reexamined in order to make it
applicable to the case of effective theories. We slightly modify the
formulation of this theorem and prove it basing on the notion of
generating functional for Green functions. This allows one to trace
(directly in terms of graphs) the mutual cancelation of different
groups of contributions.
\end{abstract}


\section{Introduction}
\label{Introduction}
\mbox{}

The equivalence theorem (ET) is known since the early sixties
\cite{Chisholm} -- \cite{SalamStrathdee}.
Then, from time to time it was considered by different authors
from various points of view (see, e.g.,
\cite{Lam} -- \cite{Tyutin}).
However, many practical aspects of this theorem remain unclear. In
particular, the possibility to use it in the framework of effective
theories%
\footnote{We call the quantum field theory `effective' if the
interaction Hamiltonian in the interaction picture contains
{\em all}
the types of monomials consistent with a given algebraic (linear
homogeneous) symmetry (see
\cite{WeinbEFT}, \cite{AVVV1} -- \cite{KSAVVV2}).}
still seems questionable. The point is that the very formulation of
the equivalence theorem needs a refinement in order to adjust it to
the case of effective theory.

Perhaps the most comprehensive consideration of this theorem along
with the discussion of shortcomings in the previous proofs has been
done in the papers
\cite{KalloshTyutin}, \cite{Tyutin}.
In those papers the ET is treated as the statement that
$S$-matrix
in quantum theory does not depend on the choice of variables in the
corresponding classical Lagrangian. In other words, the canonical
transformations in classical theory are considered as a kind of
symmetry and the problem of the proof of ET thus reduces to that of
conserving this symmetry in the process of quantization. Clearly, in
this approach the form of classical Lagrangian cannot be arbitrary:
the number of field derivatives must be at least finite. Moreover, the
quantum theory under consideration must be renormalizable, otherwise
the
$S$-matrix
would make no sense (both these points are specially stressed in
\cite{KalloshTyutin}).

Meanwhile, the construction of effective theory is not based on any
classical Lagrangian (or Hamiltonian), it is only restricted by the
postulated structure of asymptotic states and the requirements of
certain linear symmetry along with general principles like unitarity,
causality (cluster decomposition) and Lorentz invariance of the
$S$-matrix%
\footnote{As it has been pointed out in
\cite{WeinbEFT},
this construction itself has no other physical content beyond the
aforementioned general principles. In papers
\cite{AVVV1} -- \cite{KSAVVV2}
it is shown that such a content results from additional physical and
mathematical requirements: localizability, summability, uniformity
and reasonable asymptotic behavior of amplitudes.}.
As a consequence, the number (as well as the degree) of field
derivatives appearing in the effective theory Hamiltonian is actually
infinite. This makes the quantum theory renormalizable
{\em ab initio}
(see, e.g.,
\cite{WeinMONO})
but, at the same time, makes it impossible to point out the relevant
classical Hamiltonian construction subjected to the quantization
procedure. This means that the commonly accepted formulation (and,
hence, proofs) of ET cannot be considered suitable in the case of
effective theory. In this case one needs to prove a similar theorem
(below we call it as
{\it modified equivalence theorem}
-- MET)
stating the perturbative equivalence of
$S$-matrices
in two inherently quantum theories with different Hamiltonians.
Those Hamiltonians must be constructed from the same set of free field
operators%
\footnote{See, however, the note in the last paragraph of
Sec.~\ref{conclusion}}
and connected with one another by certain transformation of the kind
\begin{equation}
\phi(x) \rightarrow \phi(x) + \alpha \sum_{k=2}^{\infty} \int\!\!
\frac{1}{k!}\, F_k(x|y_1,...,y_k)\,
\phi(y_{\sss 1}) \ldots \phi(y_k)\, dy_{\sss 1} \ldots dy_k\, .
\label{substitution}
\end{equation}
Here
$\alpha$
is just an auxiliary parameter and, by construction,
$F_n(x|y_{\sss 1},...,y_k)$
is implied symmetric in arguments
$y_1,...,y_k$.
There is no need in refereing to quantization of
any particular classical theory. Note that usually the equivalence
theorem is formulated for local transformations or, the same, for the
case when
$$
F_k(x|y_{\sss 1},...,y_k) \sim \prod_{i=1}^{k} \delta(x-y_i)\, .
$$

Here it is pertinent to note that the term
``perturbative equivalence''
makes no sense until the perturbation schemes for both theories are
specified. Then, to compare the renormalized
$S$-matrices
in two theories one needs to perform the renormalization which, in
turn, requires fixing the relevant renormalization prescriptions. At
this point one meets a difficulty: the number of necessary
prescriptions in two theories may prove to be different, at least, at
first glance. Such a situation occurs, for example, when one performs
the nonlinear change of field variables, say,
$$
\phi \rightarrow F(\phi) = \phi + \alpha {\phi}^2
$$
in the free field Lagrangian. The resulting Lagrangian contains the
interaction term of the form
${\alpha}^2(\phi {\partial}_{\mu}\phi)^2$
and thus belongs to the class of non-renormalizable theories which
require an infinite set of counterterms and the corresponding number
of renormalization prescriptions. In fact, however, this is only true
with respect to Green functions. It can be shown that the
$S$-matrix
in such a transformed theory remains trivial. Of course, the Green
functions make no sense after removing the regularization.

The written above shows that, when formulating and proving the MET
for the case of effective theory, one needs to trace the fine mutual
cancellation of the contributions
$ \sim \alpha $
from all
$S$-matrix
graphs with a given number of external lines. There is no need in
performing the complete renormalization -- it is quite sufficient to
prove this cancellation on the regularized
$S$-matrix
graphs (those with external lines on the mass shell). Surely, the
proof must be based on the graph language. Attracting the functional
integral technique is undesirable because in the case of effective
theory the very definition of functional integral looks unconvincing:
there is no classical action needed to construct it. Alternative
definition -- the formal sum of the loop perturbation series -- also
looks unacceptable because an infinite set of graphs with the same
number of loops requires special ordering to avoid divergencies. Until
this is done even the individual terms of loop series make no sense
and, hence, their formal sum cannot be reasonably defined.

The paper is organized as follows. In
Sec.~\ref{sec_Prelim}
we explain our notations and give a list of formulae used throughout
the paper. In
Sec.~\ref{sec_formulating}
we give the precise formulation of MET statement.
Sec.~\ref{sec_proof}
is devoted to the proof of MET. It consists of 6 subsections. In
Subsec.~\ref{subsec_step1}
we write down the generating functional
$\wtd{G}$
for Green functions of the transformed theory and show that it is
connected with the functional
$G$
of the initial one as follows:
$\wtd{G} = Q G$.
Here
$Q$
is the variational operator functional fixed completely by the form
of substitution law
(\ref{substitution}).
Besides, it is shown that this operator takes the form of product of
two series: determinant and exponential. In
Subsec.~\ref{subsec_graph_technique}
we consider the simplified example which helps one to understand the
graphic technique needed to study the operator
$Q$
structure. Then we list the set of graphic rules adjusted for the
case of general transformation law
(\ref{substitution}).
In
Subsec.~\ref{subsec_general_structure}
we classify the types of graphs presenting the exponential series.
Subsec.~\ref{subsec_determinant}
is devoted to the analysis of graphic structure of determinant series.
In
Subsec.~\ref{subsec_Q_series}
it is shown that only one type (of three) of graphs survive in the
series that presents the operator
$Q$.
Subsec.~\ref{Z_new}
completes the proof. Here we calculate the field strength
renormalization constant
$\wtd{z}$
in transformed theory and show the absence of mass shift and tadpoles.
Then we calculate renormalized
$n$-point
$S$-matrix
element and demonstrate that it turns out the same as that in initial
theory.
Subsec.~\ref{conclusion}
contains the concluding remarks.

The variational functional formalism which we rely upon in this paper
is not widely known. For this reason we find it pertinent to give a
brief outlook of this formalism. This is done in
Appendix A.
Besides, when analyzing the structure of the operator
$Q$
we make use of the first Mayer's theorem
\cite{Mayer}.
The legality of this step is shown in
Appendix B.

One note is in order. From the very beginning we work in the framework
of renormalized perturbation scheme. Nevertheless, it is not difficult
to check that the correctness of the result does not depend of this
circumstance.


\section{Preliminaries}
\label{sec_Prelim}
\mbox{}

In accordance with LSZ formula the renormalized
$n$-particle
$S$-matrix
element can be obtained in four steps. First, one has to calculate the
relevant Green function
$G_n(x_1, \ldots ,x_n)$.
Second, every external line should be dotted by
$D^{-1}$,
where
\begin{equation}
D(x)= \frac{1}{(2{\pi})^4}\, \int\!\! dq\; {\rm exp}\{iqx\}\, D(q)\, ,
\ \ \ \ \ \ \ \ \ \ D(q)= \frac{i}{q^2-m^2+i0}\, ,
\label{propagator}
\end{equation}
and
$D^{-1}$ --
stands for the corresponding inverse operator%
\footnote{In the case when the line in question corresponds to a
particle with spin
$J\neq 0$
it is necessary to take account of the relevant wave function.}.
Third, every external line must be dotted by the factor
$z^{\sss -1/2}$
where
$z$
is the field strength renormalization constant
\begin{equation}
z = {\rm Res}\bigg|_{p^2=m^2} [-iG_2(p^2)] =
\left(1- \Sigma' (p^2)\bigg|_{p^2=m^2}\right)^{-1}
\label{z}
\end{equation}
(here
$G_2$
stands for the 2-point Green function while
$
\Sigma' (p^2)
$
-- for the self energy derivative with respect to
$p^2$).
At last, in the obtained expression one has to perform a transition to
the mass shell.

So, only two last steps are connected with transition to the mass
shell. For this reason on the first stage of our proof we concentrate
solely on consideration of Green functions of transformed theory. The
properties of renormalized
$S$-matrix
are studied on the second stage.

Before formulating and proving MET it is necessary to explain the
notations used below. Throughout the paper we follow the monograph
\cite{Vasiliev1}
and use shortened (matrix) notations omitting the integration symbols.
For example, the expression
$aDa$
should be understood as follows:
$$
aDa \eqdef \int\!\! dxdy\, a(x)D(x-y)a(y).
$$
Similarly, the operator expression
$
\dd{a}D\dd{a}
$
means
$$
\dd{a}D\dd{a} \eqdef
\int\!\! dxdy
\frac{\delta}{\delta a(x)}D(x-y)\frac{\delta}{\delta a(y)}\, .
$$
Here the symbol
\begin{equation}
\dd{a} \eqdef \ddx{a} \eqdef \ora{a}
\label{left_deriv}
\end{equation}
stands for the conventional left variational derivative acting to the
right (`L-derivative'). As usually,
\begin{equation}
Tr\,\, ln [1-{f'}(a)] \eqdef
-\int\!\! dx {f'}(x;x|a) - \frac{1}{2}
\int\!\! dxdy {f'}(x;y|a){f'}(y;x|a)
- \ldots\, .
\label{trace_ln}
\end{equation}
Hereafter the notation
$f^{(n)}(x;y_{\sss 1}, \ldots , y_n|a)$
is used for the functional derivative of
$f(x|a)$
with respect to
$a(z)$:
\begin{equation}
f^{(n)} \equiv f^{(n)}(x;z_{\sss 1}, \ldots ,z_n|a) \eqdef
\frac{\delta^n f(x|a)}{\delta a(z_{\sss 1}) \ldots \delta a(z_n)}\, ,\
\ \ \ \ f'\equiv f^{(1)}.
\label{funct_deriv}
\end{equation}
In what follows we call the first argument as the main index (or, the
same, ``main argument'') while
($z_{\sss 1}, \ldots, z_n$)
placed between the semicolon and
vertical line -- as ``induced''.

The generating functional of Green functions
$G_n(x_1, \ldots ,x_n)$
is defined as follows:
\begin{equation}
G(a) \eqdef \sum_{n=0} \frac{1}{n!}
\int\!\! dx_1 \ldots dx_n G_n(\ldots) a(x_1) \ldots a(x_n)\, .
\label{G-functional}
\end{equation}

Throughout the paper the notation
$\phi(x)$
is used solely for
{\it free quantum field}.
The Latin letters are used for arbitrary classical fields.

We introduce two more variational differentiation operators in
addition to the L-derivative defined by
(\ref{left_deriv}).
The R-derivative
$\ola{a}$
(acts from right to left) and LR-derivative
$\olra{a}$
(the sum of L- and R- derivatives) are defined as follows:
\begin{equation}
 f(a) \ola{a} g(a) \eqdef f'(a)g(a)\, ,
\label{R_derivative}
\end{equation}
\begin{equation}
f(a) \olra{a} g(a)  \eqdef f'(a)g(a) + f(a)g'(a).
\label{LR_derivative}
\end{equation}
It is easy to show that
\begin{equation}
\left[ \olra{a}, a \right]_- = 0;\ \ \ \ \ \ \ \ \ \
\left[ \olra{a}, \ora{a} \right]_- = 0\, .
\label{comm1}
\end{equation}

So, to calculate the
$S$-matrix
elements one needs to calculate first the generating functional for
Green functions
(\ref{G-functional})
of the theory under consideration. This can be done with the help of
relation (see
\cite{Vasiliev1}):
\begin{equation}
G(a) = \exp \left[V\! \left( \ora{a} \right) \right]
       \exp \left[ \frac{1}{2}aDa \right] \cdot 1\, .
\label{G-functional_form}
\end{equation}
Here the symbol
$\cdot 1$
is used to stress that the derivatives only operate on the second
exponential%
\footnote{Below we will also consider the
{\em variational functionals} --
those without
$\cdot 1$.}:
$\ora{a} \cdot 1 = 0.$
The variational functional
\begin{equation}
V(\ora{a}) = V(b)\bigg|_{b=\ora{a}}
\eqdef
-i\int\! dx\, H_{int}^{\rm res}(b)\bigg|_{b=\ora{a}}
\label{eff_action}
\end{equation}
stands for the
{\it resultant functional image of symmetrized form}
(Sym-form)
{\it of quantum interaction Hamiltonian}%
\footnote{In
\cite{Vasiliev1}
this construction was called as effective interaction. We prefer to
call it as `resultant image' just because modern language assigns
different meaning to the term `effective interaction'.}
$
H_{int}^{\rm res}(b)\, .
$
The structure of this object is explained in
\cite{Vasiliev1}
(brief explanations are also given in Appendix~A below). In what
follows there is no need to specify the construction of this
functional.


\section{Formulating the MET}
\label{sec_formulating}
\mbox{}

In accordance with what is written in the previous Section (and in
Appendix A) we work with the resultant image
$V(a)$
that depends on arbitrary classical source field
$a(x)$
(and with the corresponding variational functional
$V(\dd{a})$).

The MET statement is formulated as follows.
{\it Two quantum field theories
with the resultant images}%
\footnote{It is implied that both
$V$ and $\wtd{V}$
are presented by finite or formal infinite series in field
$a(x)$
and its derivatives. Also, we imply that there exists certain
regularization scheme suitable for both theories.}
$
V(a)
$
{\it and}
$\wtd{V}(a)$
{\it lead to the same renormalized}
$S$-{\it matrix
under the condition that}
\begin{equation}
\wtd{V}(a) = V(a-f(a)) + a D^{-1}f(a) - \frac{1}{2}\,f(a)D^{-1}f(a)
+{\rm Tr}\,\, {\rm ln} [1-{f'}_{\sss a}(a)]\, .
\label{V1_V2}
\end{equation}
Here
$f(a)$
stands for the functional series (finite or formal infinite)
\begin{equation}
f(x|a) =  \sum_{m=2}^{\infty} f_m(x|a) =
\alpha \sum_{m=2}^{\infty} \frac{1}{m!}
\int\!\! F_m(x|y_{\sss 1},...,y_m)\,
a(y_{\sss 1}) \ldots a(y_m)\, dy_{\sss 1} \ldots dy_m\, ,
\label{f_phi_x}
\end{equation}
where
$\alpha$
is just a parameter. It is tacitly implied that the Fourier transform
of
$
F_m(x|y_1,...,y_m)
$
in
$(y_1,...,y_m)$
does not contain negative powers of momenta.

Our proof of the above-formulated theorem is built upon the comparison
of two functionals:
$G(a)$
and
$\wtd{G}(a)$
constructed in accordance with
(\ref{G-functional_form})
from
$V$
and
$\wtd{V}$,
respectively. This language allows us to trace in detail the effect of
partial cancelation between groups of graphs that appear in the
transformed theory and, at the same time, does not introduce any
difficulties compared to the functional integral language used in
\cite{KalloshTyutin}.
We would like to stress that the problem of renormalizability has
nothing to do with MET: the proof applies to regularized graphs
irrelevantly to the possibility of removing regularization.


\section{The proof}
\label{sec_proof}


\subsection{Step 1: the generating functional of transformed theory}
\label{subsec_step1}
\mbox{}

According to the relation
(\ref{G-functional_form})
the generating functional
(\ref{G-functional})
for Green functions of transformed theory reads
\begin{equation}
\begin{split}
\wtd{G}(a) =
\exp \left\{ - \frac{1}{2}
f \left( \ora{a} \right) D^{-1} f \left( \ora{a} \right)
+ V \left[ \ora{a} - f \left( \ora{a} \right) \right]
+ f \left( \ora{a} \right) D^{-1} \ora{a}  \right\}   \\
\times \exp \left\{ {\rm Tr}\, {\rm ln}
\left( 1 - f_a'  \left( \ora{a} \right) \right) \right\}
\exp\left[ \frac{1}{2} a D a \right]\cdot 1\, .
\label{step1_1}
\end{split}
\end{equation}
In this Section we will show that this expression can be rewritten as
follows
\begin{equation}
\wtd{G}(a) = \exp \left\{ {\rm Tr}\, {\rm ln}
\left( 1 - f_a'  \left( \ora{a} \right) \right) \right\}
: \exp \left[ a f(\ora{a}) \right] : G(a)
\eqdef Q \left( a, \ora{a} \right)\, G(a).
\label{step1_2}
\end{equation}
The notation
$:\ldots :$
(``anti-normal form'') should be read as follows
\begin{equation}
:\exp \left[ f \left( \ora{a} \right) a  \right]\!: \eqdef
\sum_{n=0}^{\infty}
\frac{1}{n!}f^n \left( \ora{a} \right) a^n.
\label{exp_series}
\end{equation}

The relations
(\ref{R_derivative}),
(\ref{LR_derivative})
and
(\ref{comm1})
allow one to change the direction of arrows above certain operators
$\ora{a}$
in
(\ref{step1_1})
and rearrange them as follows (recall that
$1 \cdot \ola{a} = \ora{a} \cdot 1 = 0$):
\begin{equation}
\begin{split}
\wtd{G}(a) = 1 \cdot {\cal D} \cdot  {\cal F} \cdot
\exp{ V \left[ \ora{a} - f \left( \olra{a} \right) \right]}
\exp{\left[ f \left( \olra{a} \right) D^{-1} \ora{a} \right] }
\exp\left[ \frac{1}{2} a D a \right]\cdot 1 {}\, .
\label{step1_3}
\end{split}
\end{equation}
Here
\begin{equation}
{\cal D} \eqdef  \exp \left\{ {\rm Tr}\,\, {\rm ln}
\left[ 1 - f_a'  \left( \olra{a} \right) \right] \right\} =
{\rm det}\left[ 1 - f_a'  \left( \olra{a} \right) \right]
\label{step1_4}
\end{equation}
and
\begin{equation}
{\cal F} \eqdef \exp \left\{ -\, \frac{1}{2}
f \left( \olra{a} \right) D^{-1} f \left( \olra{a} \right) \right\}.
\label{step1_5}
\end{equation}
Because
$f(\olra{a})$
commutes with
$a$
one can make use of the identity
\begin{equation}
\exp \left[ h \ora{a}\right] F(a)\cdot 1 = F(a+h)\, ,
\label{identity1}
\end{equation}
and rewrite two last exponentials in
(\ref{step1_3})
as follows:
\begin{equation}
\begin{split}
\exp & \left[ f \left( \olra{a} \right) D^{-1} \ora{a} \right]
\exp \left[ \frac{1}{2} a D a \right]\cdot 1 = \\
= & \exp \left\{ \frac{1}{2}
\left[a+f\left(\olra{a}\right)D^{-1}\right]D
\left[a+f\left(\olra{a}\right)D^{-1}\right] \right\}\cdot1 = \\
= & \exp  \left[ \frac{1}{2}
f\left(\olra{a}\right)D^{-1} f\left(\olra{a}\right) \right]
\exp \left[ a f \left( \olra{a} \right) \right]
\exp \left(\frac{1}{2}aDa \right)\cdot 1 = \\
= &\, {\cal F}^{-1} \exp \left[ a f \left( \olra{a} \right) \right]
\exp \left(\frac{1}{2}aDa \right) \cdot 1\, .
\label{step1_6}
\end{split}
\end{equation}
The last equality follows from the definition
(\ref{step1_5}).
Substituting
(\ref{step1_6})
in
(\ref{step1_3})
we obtain
\begin{equation}
\wtd{G}(a) =
1 \cdot {\cal D} \cdot
\exp{ V \left[ \ora{a} - f \left( \olra{a} \right) \right]}
\exp \left[ a f \left( \olra{a} \right) \right]
\exp \left(\frac{1}{2}aDa \right)\cdot 1\, .
\label{step1_7}
\end{equation}

Let us now make use of another identity, namely,
\begin{equation}
F\left(\ora{a}\right)\exp[ha]\cdot\Phi (a)=
\exp [ha]
F \left( \ora{a} + h \right) \cdot \Phi (a)\, .
\label{identity2}
\end{equation}
It allows one to rewrite
(\ref{step1_7})
as follows
\begin{equation}
\begin{split}
\wtd{G}(a)  = & 1 \cdot
{\rm det} \left[ 1 - f_a'  \left( \olra{a} \right) \right]
\exp \left[ a f \left( \olra{a} \right) \right]
\exp \left[ V \left( \ora{a} \right) \right]
\exp \left( \frac{1}{2}aDa \right)\cdot 1\,\, .
\label{step1_8}
\end{split}
\end{equation}

The relations
(\ref{comm1})
allow one to change the direction of arrows in
(\ref{step1_8}).
Indeed,
\begin{equation}
\begin{split}
\wtd{G}(a) =\,\, & 1 \cdot
{\rm det} \left[ 1 - f_a'  \left( \olra{a} \right) \right]
\sum_{k=1}^{\infty}\frac{1}{k!}
\left[ a f \left( \olra{a} \right) \right]^k G(a) \equiv \\
\equiv\,\, & 1 \cdot
{\rm det} \left[ 1 - f_a' \left( \olra{a} \right) \right]
\sum_{k=1}^{\infty}\frac{1}{k!}
\left[ f \left( \olra{a} \right) \right]^k a^k\,\, G(a)
\eqdef \\ \eqdef\,\, & 1 \cdot
{\rm det} \left[ 1 - f_a'  \left( \ora{a} \right) \right]
:\exp \left[ f \left( \ora{a} \right) a \right]\!: G(a)\, .
\label{step1_10}
\end{split}
\end{equation}
Here the account was taken of the relation
(\ref{G-functional_form}).
So, we obtain finally
\begin{equation}
\wtd{G}(a) =
1 \cdot {\rm det} \left[ 1 - f_a'  \left( \ora{a} \right) \right]
:\exp \left[ f \left( \ora{a} \right) a \right]\!:G(a)
\equiv  Q \left( a, \ora{a} \right)\, G(a)\, .
\label{Q_right_arrow}
\end{equation}
Now the relation
(\ref{step1_2})
is obtained and the first step is done. Recall that both the
determinant and exponential are just shortened notations for
corresponding operator series.

Our next goal is to bring the operator
$Q \left( a, \ora{a} \right)$
to the
``normal''
form where all source fields
$a(x)$
are placed in front of all variational operators
$\dd{a}$.


\subsection{Step 2: the operator graphic technique}
\label{subsec_graph_technique}
\mbox{}

Let us now turn to a consideration of the structure of operator
\begin{equation}
Q \left( a, \ora{a} \right) \eqdef
1 \cdot {\rm det} \left[ 1 - f_a'  \left( \ora{a} \right) \right]
:\exp \left[ f \left( \ora{a} \right) a \right]:\, .
\label{Q-form}
\end{equation}
This will require developing the special graphic technique which,
however, has nothing to do with conventional Feynman graphs of the
quantum theory in question.

To do this step by step we first consider the exponential series
(\ref{exp_series}).
The simple example discussed below allows one to better understand
the graphical structure of this series.

Suppose that the sum in
(\ref{f_phi_x})
contains only one term
\begin{equation}
f_m(x|a)=0, \ \ \ \ (m \neq k).
\label{example1}
\end{equation}
For simplicity we took
$\alpha = 1$.
In this case the exponential under consideration takes the form
(here
$
\vec{y}_i \eqdef (y_{i\!{\sss 1}}, \ldots, y_{ik}).
$)
\begin{multline}
:\exp \left[ f \left( \ora{a} \right) a \right]:\, \, =
\sum_{n=0}^{\infty} e_n =
\sum_{n=0}^{\infty} \frac{1}{n!} \int\!\!dx_{\!\sss 1} \ldots dx_n
d\vec{y}_{\sss 1} \ldots d\vec{y}_n
\left( \frac{1}{k!}\right)^n\, \times
\\
F_k(x_{\!\sss 1}|\vec{y}_{\sss 1}) \ldots F_k(x_n|\vec{y}_n)
\dds{a(y_{\!\sss 1\!1})} \ldots \dds{a(y_{{\!\sss 1}k)}} \ldots
\dds{a(y_{n{\!\sss 1}})} \ldots \dds{a(y_{nk})}
a(x_{\!\sss 1}) \ldots a(x_n)\, .
\label{example2}
\end{multline}
This expression should be brought into the normal form which can be
easily interpreted graphically. First of all let us reduce the first
nontrivial term to the form where the source field
$a$
stands before all differentiation operators. This gives
\begin{multline}
e_1 =
:\left[ f \left( \ora{a} \right) a \right]:\,\,
\equiv\,\,
\int\! dx_{\!\sss 1} \frac{1}{k!}\int\! d\vec{y}_{\sss 1}\,
F_k(x_{\!\sss 1}|\vec{y}_{\!\sss 1})
\dds{a(y_{\!\sss 11})} \ldots \dds{a(y_{\!{\sss 1}k})} a(x_1)\, =
\\
\int\! dx_{\!\sss 1}
\left\{
\frac{1}{k!}\,\int\! d\vec{y}_{\sss 1}
 F_k(x_{\!\sss 1}|\vec{y}_{\!\sss 1})
\left[
\sum_{i=1}^k
\delta(x_{\!\sss 1}-y_{{\!\sss 1}i})
\prod_{r=1 \atop r\neq i}^k \dds{a(y_{{\!\sss 1}r})}
\right]\,
+\, \right.  \\  \left.
a(x_{\!\sss 1})\frac{1}{k!}\int\!  d\vec{y}_{\!\sss 1}\,
F_k(x_{\!\sss 1}|\vec{y}_{\!\sss 1})
\dds{a(y_{\!\sss 11})} \ldots \dds{a(y_{{\!\sss 1}k})}
\right\}
\, \equiv\,                                \\
\int\! dx_{\!\sss 1}
\left[
f'(x_{\!\sss 1};x_{\!\sss 1}| \ora{a})\, +\,
a(x_{\!\sss 1}) f(x_{\!\sss 1}|\ora{a})
\right]\, .
\label{example3}
\end{multline}
Note that the first term in the last line of
(\ref{example3})
results from the integrating with relevant
$\delta$-functions
and accounting for the symmetry of
$F_k(x|y_{\sss 1}, \ldots ,y_k)$
in arguments
$(y_{\sss 1}, \ldots ,y_k)$.
Note, also, that in this term the induced argument coincides with the
main one.

To better understand the structure of higher terms of the series
(\ref{exp_series})
let us calculate the third term
$$
e_2 = \frac{1}{2!}\int\!\!dx_{\sss 1} dx_{\sss 2}
d\vec{y_{\sss 1}} d\vec{y_{\sss 2}}
F_2(x_{\sss 1}| \vec{y_{\sss 1}}) F_2(x_{\sss 2}| \vec{y_{\sss 2}})
\dds{a(y_{\sss 11})}\dds{a(y_{\sss 12})}
\dds{a(y_{\sss 21})}\dds{a(y_{\sss 22})}
a(x_{\sss 1}) a(x_{\sss 2})
$$
of the expansion
(\ref{example2})
using the compact notations
(\ref{funct_deriv})
(for brevity we consider the case
$k=2$).

This gives (the integrating over repeating arguments
$x_1, x_2$
is implied; the operator arguments
$\ora{a}$
are not shown):
\begin{multline}
e_2 = \frac{1}{2!} f_2(x_1|)f_2(x_2|)a(x_1)a(x_2) =
\frac{1}{2!}f_{\sss 2}(x_1|)
\left[ a(x_1)f_2(x_2|)+f_2^{(1)}(x_2;x_1|) \right]a(x_2)
= \ldots = \\
\frac{1}{2!}
\left\{
a(x_1)a(x_2)f_2(x_1|)f_2(x_2|) +
a(x_2)f_2^{(1)}(x_1;x_1|)f_2(x_2|)+
a(x_1)f_2^{(1)}(x_2;x_2|)f_2(x_1|)+
\right. \\  \left.
f_2^{(1)}(x_1;x_1|)f_2^{(1)}(x_2;x_2|) +
a(x_1)f_2^{(1)}(x_1;x_2|)f_2(x_2|) +
a(x_2)f_2^{(1)}(x_2;x_1|)f_2(x_1|) +
\right. \\  \left.
f_2^{(1)}(x_1;x_2|)f_2^{(1)}(x_2;x_1|) +
f_2(x_2|)f_2^{(2)}(x_1;x_1,x_2|) +
f_2(x_1|)f_2^{(2)}(x_2;x_1,x_2|)
\right\}.\qquad
\label{compact_form}
\end{multline}
Here the relation
\begin{equation}
f^{(p)}_k(x_i;z_{i{\sss 1}}, \ldots , z_{ip}|) a(x_r) =
a(x_r)f^{(p)}_k(x_i;z_{i{\sss 1}}, \ldots , z_{ip}|) +
f^{(p+1)}_k(x_i\, ;x_r,z_{i{\sss 1}}, \ldots , z_{ip}|)
\label{commutator}
\end{equation}
has been multiply used.

The above considered examples allow one to formulate simple rules
needed to present the exponential series in graphic form. Indeed,
calculating higher terms of the expansion
(\ref{example2})
($e_{\sss 3}$, $e_{\sss 4} \ldots $)
we see that the result always takes a form of sum of products of
independent (``disconnected'') integrals over repeating arguments of
relevant factors (in operator graphic language -- vertices, or, more
precisely,
{\em vertex variational functionals}).
Let us consider the graphic structure of individual independent
integrals (``connected graphs'').
For this it is necessary to formulate the set of
rules needed to present the terms of the expansion
(\ref{exp_series})
(under the condition
(\ref{example1}))
in graphical form. This set reads:
\begin{enumerate}
\item
The
$n$th
order term is presented by the sum of oriented graphs (both connected
and disconnected). Every individual item (graph) of this sum has
$n$
vertices marked by the main arguments of relevant vertex factors.
\item
Every connected graph consists of vertices, oriented propagator lines
(propagators) connecting the vertices with one another or a given
vertex with itself, and free lines of two kinds: outgoing and
incoming. The analytical form corresponding to a given connected graph
should be constructed in accordance with the rules listed below.
\item
Every vertex has one incoming line and
$k$
outgoing ones.
\item
The vertex marked by its main index
$x_{\sss 1}$
with
$p$
outgoing propagator lines (which connect it with vertices marked by
$z_{\sss 1}, \ldots, z_p$)%
\footnote{One of
$z_i$
may coincide with
$x_{\sss 1}$.
Besides, it is clear that
$p \leq k$.},
$(k-p)$
outgoing free lines and one incoming line (the type of which is
immaterial) corresponds to the non-local variational operator
expression
$$
f_k^{(p)}(x_{\sss 1};z_{\sss 1}, \ldots, z_p|).
$$
\item
The incoming free line (of the vertex marked by
$x_r$)
corresponds to
$a(x_r)$.
\item
The propagator line is oriented: it starts at the vertex with index
$x_k$
and ends at that with index
$x_l$.
This means that one of induced arguments of the vertex marked by
$x_k$
coincides with
$x_l$.
The corresponding factor is just a unity.
\item
The integration over the main indices of all vertices is implied.
\item
Every graph of
$n$th
order should be dotted by the factor
$1/n!$
appearing in the exponential series.
\item
To avoid confusion one should arrange the product of factors
associated with elements of a given graph such that the source field
$a(x_i)$
is placed first.
\end{enumerate}

\begin{figure}[t!] 
 \begin{center}
 \includegraphics{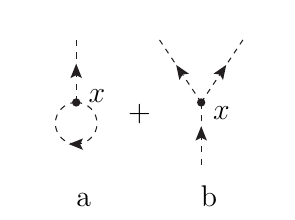}
 \caption{\label{Fig.1} The graphical representation of the terms
                        in the fourth line of
                        (\ref{example3}).}
 \end{center}
\end{figure}

\begin{figure}[t!] 
 \begin{center}
 \includegraphics{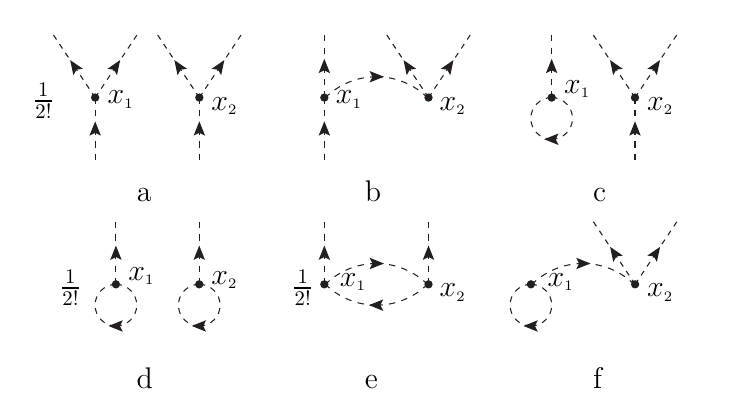}
 \caption{\label{Fig.2} The third term of the series (\ref{example2}).
                        The integral over
                        $x_{\sss 1}$ and $x_{\sss 2}$
                        is implied.}
 \end{center}
\end{figure}

The graphical representations of the expressions
(\ref{example3})
and
(\ref{compact_form})
are shown on
Figs.~\ref{Fig.1}
and
\ref{Fig.2},
respectively.

\begin{figure}[t!] 
 \begin{center}
 \includegraphics{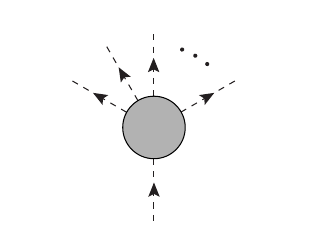}
 \caption{\label{Fig.3} Non-local composite vertex.}
 \end{center}
\end{figure}

Clearly, if the function
$f(x|a)$
is defined by the general relation
(\ref{f_phi_x}),
the corresponding complex vertex are constructed as direct sum of
elementary ones described above.  Such a composite vertex can be drawn
as a bullet (see
Fig.~\ref{Fig.3}).
From the above-listed rules it follows that: 1) Connected graph can
have at most one incoming line; 2) The connected graph cannot have
more than one loop; 3) The one-loop graphs have no incoming line or,
the same, such graphs do not contain the source field
$a(x)$.

The given above set of compact graphic rules is convenient just
because every connected graph presents nothing but a product of
relevant vertex factors (integrated over their main indices), the
factor that contains the source field
$a(x_i)$
being placed in front of all others.

It seems us pertinent to note that in often discussed case when
the function
$F_n$
is taken singular
$$
F_n(x|y_{\!\sss 1} \ldots y_n) \sim
\prod_{i=1}^n \delta(x-y_i)
$$
the loop graphs become singular and one needs to introduce
regularization. This fact is usually neglected. Nevertheless, this is
not a mistake. Below we will prove that the loop graphs stemming from
the determinant series cancel those stemming from exponential. This
means that the result does not depend on the type of regularization
chosen. In our approach the regularization appears in natural way: it
is provided by non-local character of the transformation law
(\ref{substitution}).


\subsection{The exponential series: general structure}
\label{subsec_general_structure}
\mbox{}

The discussion in previous Subsection allows one to conclude that the
exponential series
(\ref{exp_series})
results in three different types of
{\it connected}
operator graphs%
\footnote{Perhaps it makes sense to stress that individual connected
operator graphs, that appear as multipliers in the structure of
disconnected ones (see, e.g., the terms c and d of the sum
presented on
Fig.~\ref{Fig.2}),
do not interact. They only work on
$\wtd{G}(a)$.}:
\begin{itemize}
\item[\bf A.]
Trees. They have one incoming line, the others are outgoing (like the
graph
{\bf b}
in
Fig.~\ref{Fig.2}).
This means that each of these graphs contains the field
$a$
surviving after all variational derivatives inside the exponential are
done.
\item[\bf B.]
One particle irreducible (1PI) one loop graphs (``garlands''). These
graphs have no incoming lines coordinated with the source field (see,
e.g., the graph {\bf e} and every one of two subgraphs {\bf d} on
Fig.~\ref{Fig.2})
\footnote{The graph with self-closed loop (like one of subgraphs
{\bf d}
on
Fig.~\ref{Fig.2})
appears when one of the induced arguments of a given vertex factor
coincides with its main index. The graph like one of subgraphs {\bf e}
appears when the main index of factor
$V_1$
coincides with the induced argument of factor
$V_2$
and vice versa.}.
\item[\bf C.]
1PI one loop graphs connected to some number of tree graphs (e.g., the
graph
{\bf f}
of the sum presented on
Fig.~\ref{Fig.2}).
This kind graphs also have only outgoing lines.
\end{itemize}

Recall that the exponential does not produce connected graphs with
the number of loops
$l > 1$.


\subsection{The determinant series}
\label{subsec_determinant}
\mbox{}

Let us now turn to a consideration of the determinant series
\begin{equation}
{\cal D} =  \exp \left\{
{\rm Tr}\,\, {\rm ln} \left[ 1 - f_a' \left( \ora{a} \right) \right]
\right\}.
\label{det_3}
\end{equation}
As it follows from
(\ref{trace_ln})
and the listed above set of compact graphic rules, the expression
(\ref{det_3})
can be graphically presented in the form of an infinite series of
disconnected garlands. Every garland has the same vertices and is
constructed precisely in the same way as the above-described graph
of the type
{\bf B}.
The first nontrivial term in the series
(\ref{det_3})
reads
\begin{equation}
- \sum_{k=1}^{\infty}\frac{1}{k}\,\, L_k\, ,
\label{trace_ln_2}
\end{equation}
where
$L_k$
stands for the garland with
$k$
vertices.


\subsection{The operator Q series}
\label{subsec_Q_series}
\mbox{}

From the written above it follows that the operator
$Q$,
which is the product of two series -- determinant and exponential,
results precisely in the same graphs as those stemming from the
exponential; the only difference is connected with the values of
combinatorial factors. Indeed, the garlands from the determinant
series only may act on trees produced by exponential -- they commute
with exponential loop graphs just because the latter ones have no
incoming arrows.

Let us show now that, in fact, only the graphs
of the type
{\bf A}
survive in this product: the loop graphs cancel each other. For this
it is sufficient to show that the sum of
{\em connected}
graphs
$ W(\ora{a},a)$
does not contain loop contributions%
\footnote{The proof of this statement is given in Appendix B.}.

First of all, it is clear that the
$k$-vertex
graphs of the type
{\bf B}
(garlands) stemming from the determinant completely cancel those from
the exponential. This follows from the comparison of relevant symmetry
factors: as pointed above (see
(\ref{trace_ln_2}))
the symmetry factor of the determinant
garland equals
$-1/k$
while that of the exponential one is
$1/k$.
The latter value results from the product of two factors: the
numerical coefficient
$1/k!$
that appears in the exponential series and
$(k-1)!$
provided by the garland-type contribution from the operator term
$
:\!\prod_{i=1}^k f(x_i|\ora{a})a(x_i)\! :.
$
Recall that we only consider the connected graphs.

Next, let us consider the type
{\bf C}
graphs with only one
$l$-vertex
tree (``tail'') connected to the
$k$-vertex
garland. There are two different sources of such kind graphs. First,
as explained above, they appear in the exponential series. The
corresponding combinatorial factor we denote as
$S_{exp}^C(k,l)$.
Second, they appear as the result of acting the
$k$-vertex
garland from determinant on the relevant
$l$-vertex
tree from exponential (graph {\bf f} on
Fig.~\ref{Fig.2}).
The symmetry factor for the graph stemming from this latter source we
denote as
$S_{d - e}^C(k,l)$.
We need to show that
\begin{equation}
S_{exp}^C(k,l) = - S_{d - e}^C(k,l).
\label{sym_fact1}
\end{equation}

It is not difficult to show that the symmetry coefficient for
$k$-vertex
garland equals
$1/k$.
The detailed structure of the
$l$-vertex
tree graph (the number of its branches) is not essential for the
further analysis. Suppose its combinatorial factor is
$S^A(l)$.
Then the factor
$S_{exp}^C(k,l)$
reads:
\begin{equation}
S_{exp}^C(k,l) = \frac{1}{(l+k)!}\, C^k_{l+k}\, (k-1)!\, k\,  S^A(l) =
S^A(l).
\label{sym_fact2}
\end{equation}
Here the multiplier
$C^k_{l+k}$
accounts for the possibility to choose
$k$
vertices needed to construct the garland from
$(k+l)$
ones,
$1/(k+l)!$
stems from the exponential series, while the meaning of
$(k-1)!$
is explained above. At last, the factor
$k$
accounts for the possibility to choose one of
$k$
vertices of the garland which the tree is connected to.

Let us now compute the factor
$S_{d - e}^C(k,l)$.
Taking account of
(\ref{trace_ln_2})
we obtain
\begin{equation}
S_{d - e}^C(k,l) = -\, S^A(l)\, k \frac{1}{k} = -\, S_{exp}^C(k,l).
\label{sym_fact3}
\end{equation}
Here the factor
$k$
appears due to the same reason as in
(\ref{sym_fact2})
while
$1/k$
is the symmetry coefficient of a garland. This shows that the relation
(\ref{sym_fact1})
is true in the case of one-tail garlands. The generalization to the
case of garlands with many tails is straightforward.


\subsection{Step 3: mass shift, tadpoles and the field renormalization
            constant}
\label{Z_new}
\mbox{}

\begin{figure}[t!] 
 \begin{center}
 \includegraphics{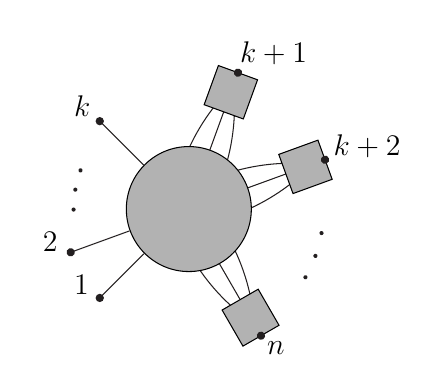}
 \caption{\label{Fig.4} The typical connected Green function graph
                          of transformed theory. It is implied that
                          the circle corresponds to the full
                          (containing both strongly and weakly
                          connected parts) connected Green function
                          of the initial theory..}
 \end{center}
\end{figure}

In this Subsection we closely follow the reasoning of
\cite{KalloshTyutin}.
As shown above, the graphs for
$\wtd{G}(a)$
result from the action of the series of
{\em operator trees}
stemming from the exponential on the graphs of generating functional
$G(a)$
of the initial theory. Therefore the typical graph%
\footnote{We only classify the connected parts of the
{\em full graphs} --
those constructed from the full lines and full vertices;
generalization is trivial.}
of the
$n$-point
Green function
$\wtd{G}_n(x_1,\ldots,x_n)$
of transformed theory appears as that presented on
Fig.~\ref{Fig.4}.
On that Figure the circle corresponds to the connected (amputated)
Green function of the initial theory while squares present the full
sum of operator trees (the type
{\bf A}
graphs described in
Sec.~\ref{sec_proof}).
It is implied summation over the number of operator subgraph legs.

\begin{figure}[t!] 
 \begin{center}
 \includegraphics{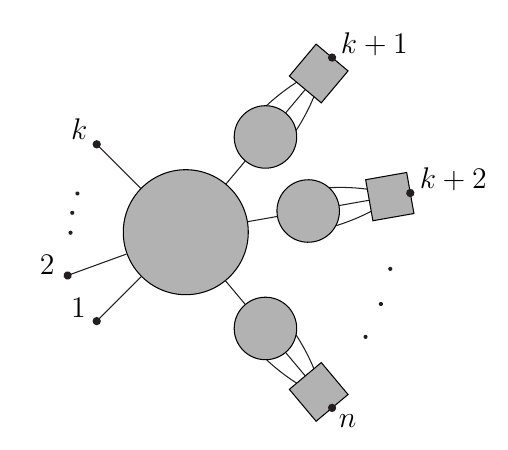}
 \caption{\label{Fig.5} Green function graph with $\sigma$-legs.}
 \end{center}
\end{figure}

It is clear that the only type of Green function graphs
providing nontrivial contribution to
$n$-particle
$S$-matrix
elements of transformed theory is that presented on
Fig.~\ref{Fig.5}
with
$k=0,1,\ldots, n$.
These graphs have two kinds of legs:
\begin{itemize}
\item[\bf A]
`Conventional' legs each of which corresponds to the full propagator
of initial theory (on
Fig.~\ref{Fig.4}
they are marked by numbers
$1,2,\ldots,k$).
\item[\bf B]
The so-called
$\sigma$-legs
(marked by numbers
$k+1,k+2,\ldots,n$
on
Fig.~\ref{Fig.4}).
Every one of such legs presents the 2-point
$\sigma$-graph
(shown on
Fig.~\ref{Fig.6})
weakly connected to the rest part of the whole graph in question.
{\em We
would like to stress that, by construction,
$\sigma$-graphs
have no poles in}
$p^2$.
\begin{figure}[t!] 
 \begin{center}
 \includegraphics{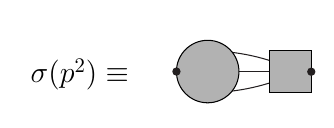}
 \caption{\label{Fig.6} The structure of $\sigma$-graph. By
                       construction, it has no pole in $p^2$.}
 \end{center}
\end{figure}
\end{itemize}

\begin{figure}[t!] 
 \begin{center}
 \includegraphics{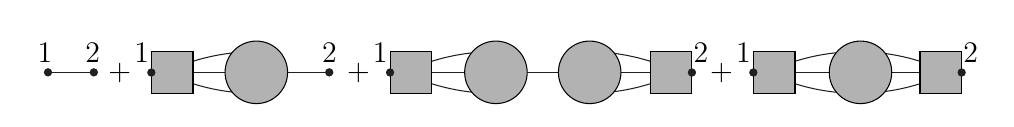}
 \caption{\label{Fig.7} Full 2-point Green function $\wtd{G_2}(p^2)$.}
 \end{center}
\end{figure}

To calculate the renormalized
$S$-matrix
in transformed theory we need first to consider two point Green
function
$\wtd{G}_2$
in order to find its pole position and the corresponding residue.
The relevant sum of graphs is shown on the
Fig.~\ref{Fig.7}.
Recall that every solid line corresponds to the full propagator of the
initial theory and all vertices are implied full. In other words,
every graph depicted in
Fig.~\ref{Fig.7}
presents an infinite series. Note that the last graph in this sum is
one particle irreducible.

Let us now suppose that the pole of full two point Green function of
the initial theory is located at
$p^2=m^2$.
From
Fig.~\ref{Fig.7}
it follows immediately that the pole position of two point function
$\wtd{G}_2(p^2)$
of transformed theory remains unchanged -- the mass shift is absent.

Similar analysis shows that the tadpole graphs do not appear in the
transformed theory if -- as we tacitly imply -- they were absent (due
to relevant renormalization condition) in the initial one.

Further, denoting the
$\sigma$-graph
(Fig.~\ref{Fig.6})
at
$p^2=m^2$
as
$\sigma$
one can present the expression for the residue of the full propagator
of transformed theory (see
Fig.~\ref{Fig.7};
note that only the first three graphs contribute, the second one
having the symmetry factor 2) as follows:
\begin{equation}
{\rm Res}\, \wtd{G}_2 \bigg|_{p^2=m^2} =
i\left[ 1 + 2 \sigma + {\sigma}^2 \right]\cdot
{\rm Res}\, G_2 \bigg|_{p^2=m^2}\, .
\end{equation}
Therefore the field strength renormalization constant in the
transformed theory differs from that in the initial one; it reads:
\begin{equation}
\wtd{z} = -i\, {\rm Res}\, \wtd{G}_2 \bigg|_{p^2=m^2} =
z\, [1 + \sigma ]^2\, .
\label{new_z}
\end{equation}

Let us now consider the new renormalized
$S$-matrix.
The connected
$n$-leg
$S$-matrix
graph in the transformed theory can be obtained from the
$n$-point
Green function graph
$
\wtd{G}_n
$
with the help of substitution
\begin{equation}
a_i \longrightarrow \wtd{z}^{-1/2} D^{-1}(p_i)
\end{equation}
and subsequent taking the limit
$p_i^2 \rightarrow m^2.$
This results in disappearance of factors
$D$
from external lines and from
$\sigma$-legs.
Besides, this introduces the common factor
$\wtd{z}^{-n/2}.$
As is seen from
Fig.~\ref{Fig.5},
the resulting factor that appears after summing graphs with different
number of
$\sigma$-legs
reads:
\begin{equation}
\wtd{N}=\wtd{z}^{-n/2} \sum_{k=0}^n C_n^k \sigma^k =
[1+\sigma ]^n z^{-n/2}[1+\sigma]^{-n} = z^{-n/2}\, .
\end{equation}
Here the multiplier
$C_n^k$
accounts for various possibilities to choose the
$(n-k)$
$\sigma$-legs
from the total number
$n$
of legs of the graph in question.

Thus it is shown that (the symbol
$\wtd{G}_n(k)$
stands for
$n$-point
Green function with
$k$
$\sigma$-legs)
\begin{equation}
\wtd{z}^{-n/2} \sum_{k=0}^n \wtd{G}_n(k) = z^{-n/2}G_n .
\end{equation}
From this it follows immediately that the renormalized
$n$-leg
$S$-matrix
graphs in two theories (initial and transformed) coincide identically.
In other words, these two theories result in the same renormalized
$S$-matrix:
$$
\wtd{S}^{\rm ren} = S^{\rm ren}
$$
This proves the theorem under consideration.


\section{Concluding remarks}
\label{conclusion}
\mbox{}

As we have already mentioned in
Sec~\ref{Introduction},
the very formulation of the proved above theorem differs from that
considered previously in the literature (see, e.g.,
\cite{KalloshTyutin}, \cite{Tyutin}).
For this reason it makes sense to discuss the consequences in more
detail.
\begin{enumerate}
\item
Our proof is not connected with any quantization scheme. It is not
connected with the structure of classical field theory at all. We work
directly with the quantum interaction Hamiltonian in the interaction
picture and rely upon the conventional perturbation scheme (Dyson's
$T$-exponential
presented in the form of generating functional for Green functions).
This makes the MET applicable for the case of effective theory%
\footnote{The problem of the number and concrete form of required
renormalization conditions lies beyond the scope of MET. In any case
the correctness of the above-given proof is not based on any
suggestions concerning this point. Also, the proof is equally
applicable to the case of spin-1/2 fields. The generalization to
the case of higher spin fields requires certain complications.}.
\item
Our treatment -- in contrast to the conventional one -- is well
suited for a consideration of
``good enough''
non-local field transformations. This is especially important in
effective theories which are non-local by their very construction.
\item
The above-given proof does not require using the functional integral
technique which looks unconvincing in the case of effective theory.
For this reason our result may be considered as independent proof of
the admissibility of changing variables in the formally written
functional integral that depends on the special kind of source
function.
\item
It is pertinent to note that the substitution of the form
(\ref{f_phi_x})
adds an additional set of parameters to that presented in the initial
theory. For this reason when comparing two
$S$-matrices
one needs to work within the perturbation scheme defined by the
initial theory. In this case, as it follows from our proof, all the
terms depending on new parameters, will disappear at every order of
the initial perturbation expansion.
\item
Clearly, our proof can be easily generalized for the case of infinite
number of scalar (and spinor) fields.
\end{enumerate}

At last we would like to mention the idea first suggested by Veltman
\cite{Veltman}.
In that paper it was shown that the local (very simple) theory of one
stable and one unstable particle with masses
$m$ and $M > 2m$,
respectively, can be reformulated solely in terms of the stable
particle field with non-local Lagrangian. The resulting
$S$-matrix,
defined on the space of asymptotic states of stable particles, turns
out to be unitary and causal; its elements are the same as the
corresponding
$S$-matrix
elements in initial (simple and local) theory. Does it mean that one
can find, at least in principle, the equivalent transformation of a
local theory with stable particle
$\phi$
and resonance
$\kappa$
to the non-local one only containing the
$\phi$
field? It seems us interesting to explore such a possibility. The MET
may turn out to be a useful tool for this purpose.


\section*{Acknowledgements}
\label{Acknowledgements}
\mbox{}

We dedicate this paper to the memory of our Teacher -- professor
A.~N.~Vasiliev. His excellent monograph
\cite{Vasiliev1}
served us as the desk-top book in the process of our work on the proof
of MET.

We would like to express our sincere gratitude to A.~Tochin for
valuable contribution to our understanding of the problems discussed
in this paper and for his help in making Figures. Also we are grateful
to V.~Blinov, M.~A.~Braun, V.~A.~Franke, S.~Paston, Ju.~Pis'mak and
M.~Vyazovsky for stimulating discussions. The work of D.~Chicherin was
supported by the Chebyshev Laboratory (Department of Mathematics and
Mechanics, Saint-Petersburg State University) under the grant
11.G34.31.0026 of the Government of Russian Federation.


\section*{Appendix A}
\label{App_A}
\mbox{}

Since the variational functional technique described in the monograph
\cite{Vasiliev1}
is not as widely known as, say, the functional integral, it seems
pertinent to devote this Appendix to brief outline of certain aspects
which we rely upon in the main text.

First we have to remind what is Sym-form. The precise definition of
Sym-form of operator product looks as follows:
$$
{\rm Sym}\left[ Q_1Q_2 \ldots Q_n \right] \eqdef
\frac{1}{n!}\sum_P \varepsilon_{\sss P}
P \left[ Q_1Q_2 \ldots Q_n \right].
$$
Here summation runs over all
$n!$
permutations
$P$
of the operators
$Q_i$
while
$\varepsilon_{\sss P}$
is nothing but conventional statistical factor (-1 for permutation of
two fermionic operators and +1 for bosonic ones). The above definition
along with many useful relations connecting Sym-product with T-
(time-ordered) and N- (normal-ordered) products can be found in
\cite{Vasiliev1}.
The usefulness of using the Sym-form of quantum Hamiltonian follows
from the possibility to exploit the variational form of Wick theorems
first suggested in
\cite{Hori}.

Let us now outline the relation between two methods -- quantum and
variational classical -- of calculating Green functions. Let
$H(\phi)$
be the quantum interaction Hamiltonian (depending on free quantum
field
$\phi$
and its derivatives and containing all necessary counterterms). The
well known Dyson T-exponential together with Wick theorems allow one
to establish Feynman rules and to calculate Green functions%
\footnote{There is a fine point in this approach. In the case when the
interaction Hamiltonian contains time derivatives one has to take
care of the correct form of corresponding propagators.}. This can be
done in two ways. The first way is to rely upon quantum commutation
relations and use Wick theorems in their original form. The second
one consists of rewriting those theorems in terms of classical
(commuting or, in case of fermions, anticommuting) objects (Hori's
theorems
\cite{Hori})
and subsequent reformulating the formalism of Green function
calculations. It is this latter way which we briefly describe below.

The variational language allows one to perform the calculation of
Green functions operating with classical objects. For this it is
necessary to present
$H(\phi)$
in Sym-form. Suppose this is already done:
\begin{equation}
H(\phi) \eqdef H^{\rm Sym}_{int}(\phi) = {\rm Sym}H(\phi).
\label{H_quant_sym}
\end{equation}
To make use of the relations
(\ref{G-functional_form})
and
(\ref{eff_action})
one needs to construct the resultant image
$
H_{int}^{\rm res}(b)
$
of
$
H^{\rm Sym}_{int}(\phi).
$
This can be done as follows. First, define the time derivatives of
free quantum field
$\phi$
as new independent quantum fields
\begin{equation}
\frac{{\partial}^k \phi}{{\partial t}^k} \eqdef {\phi}_k\, .
\end{equation}
Second, calculate the relevant Dyson and Wick contractions
$D_{mn}^{\sss D}$
and
$D_{mn}^{\sss W}$
and corresponding differences
$$
\alpha_{mn} \equiv D_{mn}^{\rm {\sss D}} - D_{mn}^{\rm {\sss W}}\, ;
$$
(surely,
$
\alpha_{\sss 00}=0,
$
$
D_{\sss 00}^{\rm {\sss D}}=D_{\sss 00}^{\rm {\sss W}}=D;
$
$D(x)$
is defined in
(\ref{propagator})).
Third, rewrite the quantum interaction Hamiltonian
$
(\ref{H_quant_sym})
$
in terms of new fields
${\phi}_k$.
By construction
$$
H_{int}^{\rm  Sym}({\phi}_{0}, {\phi}_{1}, \ldots) \eqdef
H(\phi)\bigg|_{\partial^k \phi / \partial t^k = {\phi}_k} =
{\rm Sym}\,  H^{\rm Sym}_{int}({\phi}_{0}, {\phi}_{1}, \ldots)\, .
$$
Fourth, construct the
{\em classical image}%
\footnote{Recall that the notations
$a_k$
and
$b_i$
are used for arbitrary classical source fields.}
$$
H_{int}^{\sss {\rm  Cl}}(b_{\sss 0}, b_{\sss 1}, \ldots) \eqdef
H_{int}^{\rm  Sym}({\phi}_{0}, {\phi}_{1}, \ldots)
\bigg|_{{\phi}_i \rightarrow b_i}
$$
of this latter Sym-form. At last, calculate the
{\em resultant classical image}
$
H_{int}^{\rm res}(a_0,a_1, \ldots)
$
according to the relation:
\begin{equation}
{\rm exp}
\left\{ i H_{int}^{\rm res}(b_{\sss 0},b_{\sss 1}, \ldots) \right\} =
{\rm exp} \left( \frac{1}{2} \sum_{mn}
\frac{\delta}{\delta b_m}\alpha_{mn} \frac{\delta}{\delta b_n}\right)
{\rm exp}
\left\{ i  H^{\sss {\rm Cl}}_{int}(b_{\sss 0},b_{\sss 1}, \ldots )
\right\}\, \cdot 1\,
\label{res_im_1-...}
\end{equation}
and construct the resultant functional
\begin{equation}
V(a) \eqdef
-i \int \! dx\, H_{int}^{\rm res}(b_{\sss 0},b_{\sss 1}, \ldots)
\bigg|_{b_{\sss k} \rightarrow \partial^k a/ \partial t^k}\, .
\label{res_V_12...}
\end{equation}

As shown in
\cite{Vasiliev1},
the relation
(\ref{G-functional_form})
with
$V$
defined by
(\ref{res_V_12...})
results in generating functional
(\ref{G-functional})
for Green functions of the theory containing time derivatives. We
would like to emphasize that
{\em the resultant functional should be treated as just some classical
image of the given quantum interaction; it does not present the
characteristic of the classical system whose canonical quantization
could result in the quantum theory under consideration}.
We consider the quantum interaction Hamiltonian in the interaction
picture as a starting point without any refereing to the corresponding
classical system. It is this approach (first suggested by S.~Weinberg;
see
\cite{WeinMONO}
and references therein) which makes it possible to formulate the
concept of effective theory.

One more note is in order. The Sym-form of quantum interaction
Hamiltonian
$$
H(\phi_{\sss 0},\phi_{\sss 1}, \ldots)
\equiv {\rm Sym} H(\phi_{\sss 0},\phi_{\sss 1}, \ldots)
$$
depending on free quantum field
$\phi$
(and its derivatives) can be uniquely restored from its
resultant image
$H_{int}^{\rm res}(a_{\sss 0},a_{\sss 1},\ldots)$.
In turn, this Sym-form can be further rewritten in whatever form one
likes. This means that one can formulate the MET either in terms of
the classical resultant functional or in terms of Sym-form of quantum
interaction Hamiltonian. We do this in classical terms.


\section*{Appendix B}
\label{App_B}
\mbox{}

In this Appendix we would like to prove the applicability of the first
Mayer's theorem
\cite{Mayer}
for the analysis of the structure of operator
$Q(\ora{a},a)$
defined by the relation
(\ref{Q-form}).
This -- purely combinatorial -- theorem states that for arbitrary
functional
$R$
of the form
(\ref{G-functional_form})
the full sum
$S$
of its graphs can be presented as follows:
\begin{equation}
S \equiv e^{W},
\label{Mayer_theorem}
\end{equation}
where
$W$
stands for the sum of all
{\it connected}
graphs. The difficulty, of course, is that the form
(\ref{Q-form})
of the functional
$Q(\ora{a},a)$
differs from that required by Mayer's theorem -- the factor
$\cdot 1$
is absent and many terms (variational operators!) survive in addition
to those appearing in
(\ref{G-functional_form}).

To prove the desired statement let us look more closely at the
operator
$Q$
as a whole. From the written above
(see Sec.~\ref{sec_proof})
it follows that this operator
only produces the graphs of three above-described types -- no other
graphs appear in its graphical representation.

Let us consider the auxiliary functional (not an operator!)
\begin{equation}
\wtd{Q} \eqdef  1 \cdot \exp
\left\{\frac{1}{2} \sum_{i,j=1}^2
\ddx{c_i} \Delta_{ij} \ddx{c_j} \right\}
\exp \left[\Lambda(c_2) + c_1 f(c_2)\right]\cdot 1\, ,
\label{wtdQ}
\end{equation}
where
\begin{equation}
(c_1,c_2) \equiv (a,b),\ \ \ \ \
\Delta_{ij} = 1 - \delta_{ij},\ \ \ \ \
\Lambda (b) \equiv {\rm Tr}\,\, {\rm ln}\! \left[1 - f'(b)\right]\, .
\label{wtdQ_explain}
\end{equation}

The identity%
\footnote{It is implied that
$[\ora{b},h]_- = 0.$}
\begin{equation}
\exp\left[ h\ora{b} \right] f(b) =
f(b+h)\exp \left[ h\ora{b} \right]
\label{identity3}
\end{equation}
(together with specific structure of the matrix
$\Delta$)
allows one to rewrite
$\wtd{Q}$
as follows%
\footnote{Recall that, according to
(\ref{comm1}),
$a$
commutes with
$\olra{a}$.}:
\begin{equation}
\begin{split}
\wtd{Q} = & 1 \cdot  \exp\left( \olra{a} \ora{b} \right)
\exp \left[ \Lambda (b) + af(b) \right]\cdot 1 = \\
= & 1 \cdot \exp\left[ \Lambda \left(b + \olra{a} \right) \right]
\exp \left( \olra{a} \ora{b} \right)
\exp \left[ af(b) \right]\cdot 1 = \\
= & 1 \cdot \exp\left[\Lambda \left( b+\olra{a} \right)\right]
\exp\left[ af\left( b+\olra{a} \right) \right]
\exp\left[ \olra{a} \ora{b}  \right]\cdot 1\, = \\
= & 1 \cdot \exp\left[\Lambda \left( b+\olra{a} \right)\right]
\exp\left[ af\left( b+\olra{a} \right) \right]\cdot 1\, = \\
= & 1 \cdot \exp\left[\Lambda \left( b+\ora{a} \right)\right]
:\exp\left[ f\left( b+\ora{a} \right)\, a \right]\!:\cdot 1\, .
\end{split}
\label{identity3a}
\end{equation}

Since
$$
\left[b + \ora{a}, a\right]_- \cdot 1 =
\left[\ora{a}, a\right]_- \cdot 1 = I,
$$
we see that the inner lines of graphs corresponding to the functional
$\wtd{Q}$
are precisely the same as those of graphs produced by the operator
$$
Q \left( a, \ora{a} \right) \equiv 1 \cdot
{\rm det} \left[ 1 - f_a'  \left( \ora{a} \right) \right]
:\exp \left[ f \left( \ora{a} \right)\, a \right]\!:.
$$
At the same time, the presence of the factor
$\cdot 1$
in
(\ref{identity3a})
results in disappearing of the functional derivatives
$\ora{a}$
from the factors
$(b+\ora{a})$
coordinated with external lines. All this means that
the functional
$\wtd{Q}$
results in the same set of graphs and the same values of corresponding
combinatorial coefficients as the operator
$Q(a, \ora{a})$
does. To obtain the operator graphs of
$Q(a, \ora{a})$
from the functional ones of
$\wtd{Q}(a,b)$
one only needs to perform a substitution
$b \rightarrow \ora{a}$
in the factors correlated with external lines.

From
(\ref{wtdQ})
it follows that
$\wtd{Q}$
can be interpreted as the generating functional for the
$S$-matrix
of a theory of two interacting fields:
$a$
and
$b$.
The functionals of this form meet the conditions that allow one to
make use of the first Mayer's theorem.

As shown above, the operator graphs resulting from
$Q(a, \ora{a})$
are in one-to-one correspondence with the functional ones resulting
from
$\wtd{Q}$.
Therefore, it is shown that the first Mayer's theorem is quite
applicable for analyzing the graphic structure of the operator
$Q(a,\ora{a}).$

This is the statement which was to be proved.


\end{document}